\documentclass[sn-nature]{sn-jnl}

\usepackage{graphicx}%
\usepackage{multirow}%
\usepackage{amsmath,amssymb,amsfonts}%
\usepackage{amsthm}%
\usepackage{mathrsfs}%
\usepackage[title]{appendix}%
\usepackage{xcolor}%
\usepackage{textcomp}%
\usepackage{manyfoot}%
\usepackage{booktabs}%
\usepackage{algorithm}%
\usepackage{algorithmicx}%
\usepackage{algpseudocode}%
\usepackage{listings}%

\begin{document}

\title{Cross-platform impact of social media algorithmic adjustments on public discourse}


\author*[1,3]{\fnm{Pietro} \sur{Gravino}}\email{pietro.gravino@sony.com}

\author[2,3]{\fnm{Ruggiero D.} \sur{Lo Sardo}}\email{ruggiero.losardo@sony.com}

\author[2,3]{\fnm{Emanuele} \sur{Brugnoli}}\email{emanuele.brugnoli@sony.com}


\affil*[1]{\orgname{Sony CSL - Paris}, \country{France}}

\affil[2]{\orgname{Sony CSL - Rome, Joint Initiative CREF-SONY, Centro Ricerche Enrico Fermi}, \city{Rome}, \country{Italy}}

\affil[3]{\orgname{Centro Ricerche Enrico Fermi}, \city{Rome}, \country{Italy}}


\abstract{In the hypertrophic and uncharted information world, recommender systems are gatekeepers of knowledge. 
The evolution of these algorithms is usually an opaque process, but in February 2023, the recommender system of X, formerly Twitter, was altered by its chairman (Elon Musk) transparently, offering a unique study opportunity. 
This paper analyses the cross-platform impact of adjusting the platform's recommender system on public discourse. 
We focus on the account of Elon Musk and, for comparison, the account of the President of the United States (Joe Biden). 
Our results highlight how algorithm adjustments can boost content visibility, user engagement, and community involvement without increasing the engagement or involvement probabilities.
We find that higher visibility can increase the influence on social dialogue but also backfire, triggering negative community reactions. 
Finally, our analysis offers insights to detect future less transparent changes to recommender systems.}




\maketitle

\section{Introduction}\label{intro}

The advent of very large online platforms has led to a transformational shift in offline behaviour, influencing areas as diverse as the economy, social health, and democratic processes.
A growing body of academic literature has dived into this rich phenomenology, for instance, examining the role of social media in shaping public opinion on COVID-19 vaccines \cite{athey2023digital}, characterizing financial market events like the GameStop short squeeze \cite{Mancini2022}, and understanding the broader political effects of digital media on democracy \cite{zhuravskaya2020political,LorenzSpreen2022,Brugnoli2023,Mattei2022, cinelli2021echo, doi/10.2759/764631, doi:10.1177/1940161220912682}.
Recently, institutions have begun to respond with regulatory measures such as the European Union's Digital Services Act (DSA) aimed at mitigating the negative impact of digital platforms on societal well-being \cite{DSA}.

However, a key challenge to a deeper understanding of these phenomena is the lack of transparency in the operation of these platforms, an issue highlighted in the DSA. 
One of the most significant and potentially dangerous aspects of this opacity is the design and functioning of recommender systems\cite{Ricci2022,davidson2010youtube,5197422}, given their central role in almost every online platform. 
Computer science research built a substantial body of literature on improving the accuracy and efficiency of these systems \cite{Ricci2022,5197422, Silveira2017} but understanding their broader societal impact is much harder since it requires data on the evolution of the system for long time periods and the availability of information on the design and functioning of the platforms.
While the first could be available, depending on users' privacy preferences and platforms' willingness to share data with the scientific community, the latter information is typically trade secrets, with very few exceptions\cite{githubrecsys}.

Despite the difficulties, the scientific community is trying to understand the impact of recommender systems on information and opinion dynamics, which in turn is crucial for our societies. 
Measuring the impact of recommender systems beyond mere accuracy metrics is usually done through simulation, theoretical modelling, and by asking the engineers themselves~\cite{doi:10.1177/1359183518820366}. 
Several studies have pointed to systemic effects that are difficult to quantify, such as the emergence of filter bubbles~\cite{pariser2011filter}, the spread of disinformation~\cite{tommasel2022recommender, fernandez2021analysing}, increased polarization~\cite{DeMarzo2020, Sirbu2019, Baumann2020, perra2019modelling, santos2021link, whittaker2021recommender, gravino2019371, doi:10.1126/science.abp9364, Bouchaud2023} and community attempts to manipulate these systems~\cite{Si2018}. 
However, the real-time impact of algorithm adjustments remains largely invisible, typically only discernible through limited A/B testing focused on accuracy-related metrics \cite{10.1145/3159652.3159687, bottou2013counterfactual}.

The discourse around recommender systems took an interesting turn in February 2023 when entrepreneur Elon Musk openly discussed changes to the algorithm of X, i.e. the platform formerly known as Twitter. 
Since the events took place before the rebranding of the platform, in this paper we will use the old name (Twitter) and the related terminology.
This unprecedented move signals an opportunity for public scrutiny and academic investigation into the opaque world of algorithmic recommendation, offering a potential pathway to more transparent and accountable digital platforms.

At least since the beginning of February 2023, the account of Elon Musk has been used to report on platform adjustments and, in some cases, to test the platform, as reported in some tweets (e.g. to test the visibility of private versus public tweets on the 1st and 2nd of February \cite{Musk_2023a}).
Interventions on the recommendation algorithm were announced on the 9th of February \cite{Musk_2023b}, with an update in the following days (particularly the 12th \cite{Musk_2023c}). 
On the 13th of February, the recommendation algorithm was announced to be the ``top priority'' \cite{Musk_2023d}. 
The following day, ongoing adjustments to the algorithm were communicated again \cite{Musk_2023f}.
On the 14th, users started to discuss the higher presence of Musk's tweets in their personalised feeds \cite{Musk_2023e}.
On the 15th, the first report and analysis about the allegedly extraordinary changes to the algorithm of Musk's tweets started to emerge in the news \cite{platformer}.
Later the same day, \textit{\#blockelon} became the top trending hashtag on Twitter \cite{trending}.
In the following days, other reports and analyses were published \cite{washpost}.
On the 17th of February, Musk replied to newspapers and users, stating that reports were false and that the algorithm was under ongoing revisions \cite{Musk_2023h}. 
He also added that the algorithm would be published in the following weeks in an effort for transparency \cite{Musk_2023i}. 
The algorithm was published at the end of March 2023 \cite{githubrecsys}.

Reports tried to reconstruct what happened behind the scenes and why.
Musk contested these reports' versions \cite{Musk_2023g}.
Our work focuses on the impact of the adjustments to the recommendation algorithm that occurred in mid-February 2023 according to all story versions. 
Since, in some reconstructions, the President of the United States, Joe Biden, was invoked as a benchmark, we performed some analyses of the social debate about him for comparison. 
We organised our analysis focusing on three aspects: the direct impact inside the platform, the direct impact outside the platform and the indirect impact, i.e., the community's reaction. The latter was studied only inside the platform. 
Our findings highlight the effectiveness of harnessing platform algorithm control in enhancing content visibility. As a result, this can lead to a corresponding increase in user engagement and community involvement. The heightened visibility of content can, in turn, have a dual impact, influencing social discourse while also potentially inciting adverse reactions from the community.
Furthermore, our research demonstrates the value of cross-platform analysis in uncovering subtle, less transparent alterations to recommender systems that may emerge in the future.

\section{Results}\label{results}

\subsection{Internal impact on visibility and engagement}
\begin{figure}[ht]
    \centering
\includegraphics[width=.49\textwidth]{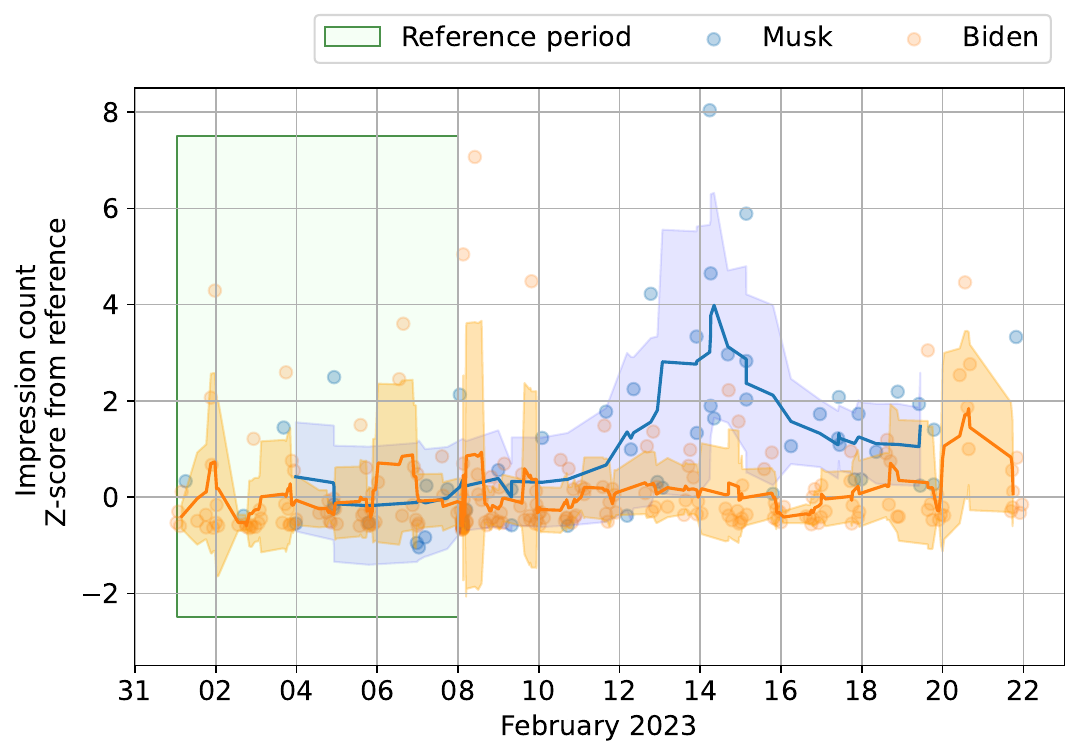}
\includegraphics[width=.49\textwidth]{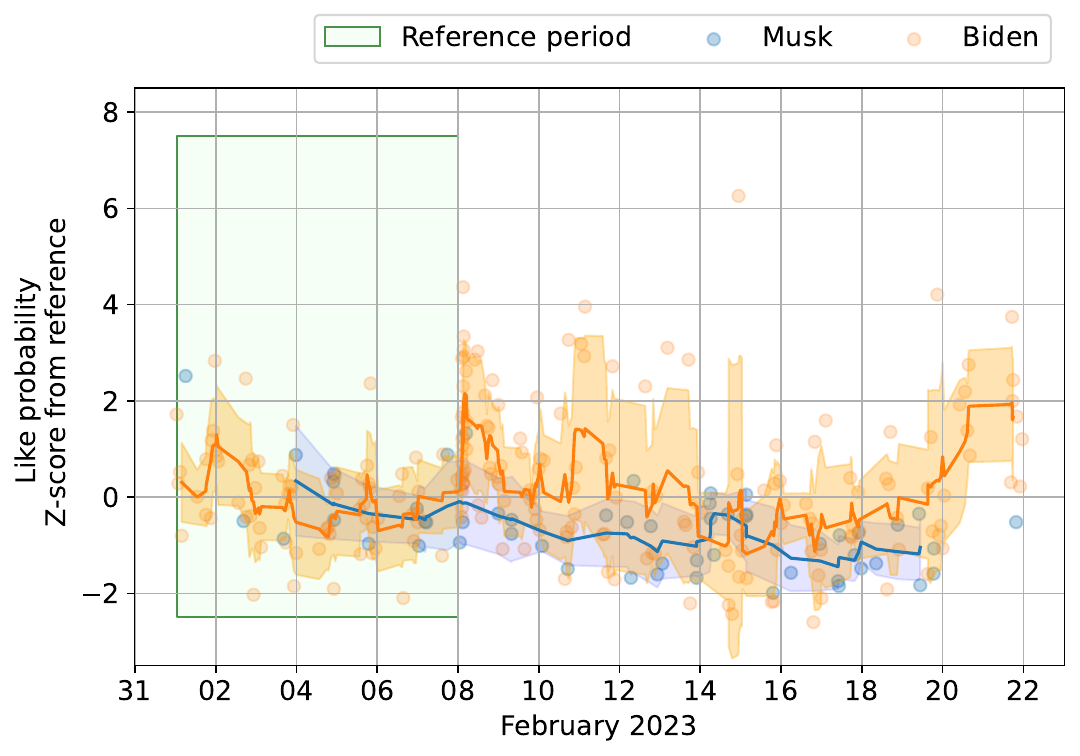}
    \caption{\textbf{Left.} The z-score of the impression count for tweets of Elon Musk and Joe Biden (benchmark). \textbf{Right.} The z-score of the like probability for Elon Musk and Joe Biden (benchmark). \textbf{Both.} The line represents the running average, and the shade represents the standard deviation (window = 7 tweets). The reference period includes the first seven days of February 2023.}
    \label{fig:intern}
\end{figure}

The main algorithmic adjustments aimed to improve the visibility of tweets. 
We assessed this by studying the metric named ``impressions'' that estimates how many times the tweet has been viewed. 
This quantity was analysed by studying the z-score calculated with respect to a period of reference (the first week of February). 
Results are reported in the left part of Fig.~\ref{fig:intern}.
From the 13th to the 16th of February, we observed an anomalous increment in the impression count of Elon Musk's time-series (with the running average hitting $z=4$). 
Such an anomaly is completely absent in the benchmark, while important outliers can still be observed punctually.

We studied the reaction triggered by these contents to see if this increase in visibility would lead to a change in the number of interactions (e.g. like, reply, etc.) or, more importantly, in the probability of interaction. 
The latter has been estimated for each tweet simply by dividing the number of interactions triggered by the tweet by the number of visualisations obtained by the tweet. 
Then, we studied the sequence of the tweets' interaction probability, calculating the z-score with respect to a reference period, similarly to what has been done in the previous paragraph. 
In the right part of Fig.~\ref{fig:intern}, we report the result for the `like' reaction. 
No systematic deviation is observable, and a similar result was observed for all the other interactions (see SI).
These results show how the increase in visibility, observed only for Musk, effectively resulted in more interactions but no increase in the probability of interaction. 

\subsection{Internal impact on community involvement}

Aside from studying the dynamics of the engagement gained by the content published by Elon Musk and Joe Biden, we also developed AutoRegressive Moving Average (ARMA) models~\cite{durbin2012time} to analyse the evolution of the communities of users interacting with such content through retweet or quote tweet.
Retweets are simple users forward of tweets on Twitter.
Quote tweets are similar but with an important difference: users comment briefly the message they forward~\cite{garimella2016quote}.
Namely, we modelled how the two profiles engaged new users over the time span analysed.
Let us first consider the quote tweet as a reference interaction.
The proposed models incorporate the reach of a tweet (i.e., the quote tweet count) and its timestamp as exogenous variables.
Indeed, we posit that the total number of quote tweets per tweet positively impacts the likelihood of new users to quote content from the same account, while the tweet timestamp negatively affects this probability compared to earlier periods (see Methods for more details).

Since we are interested in how user engagement is affected by the changes to the platform's recommendation algorithm announced by Musk on 13 February 2023, we chose this date as the watershed between train and test data for our models.
Once the models were fitted on the train data, we compared the test data with the models’ predictions to assess whether the algorithm changes coincide with an increase in the probability of interaction with content from Musk, even compared to Biden.

ARMA tools are used to model stationary stochastic processes. Hence, we first verified the time-series stationarity using the Augmented Dickey-Fuller (ADF) test~\cite{said1984}.
Subsequently, an exhaustive grid search with the Akaike Information Criterion (AIC)~\cite{akaike1974} was conducted to determine the optimal parameters for the ARMA models (see SI for details).
Results of both the tests applied to the time-series of new quoters are reported in Table~\ref{tab:ADFtest}.

\begin{table}[ht]
    \centering
    \begin{tabular}{c|cc|rr}
        & \multicolumn{2}{c|}{ADF} & \multicolumn{2}{c}{ARMA opt. pars}\\
        & Statistic & p-value & $(p,q)$ & AIC\\
        \hline
        Musk & $-8.26$ & $5.08e-13$ & $(1,2)$ & $-103.92$\\
        Biden & $-13.63$ & $1.73e-25$ & $(1,3)$ & $-205.27$\\
    \end{tabular}
    \caption{\textbf{Left.} Augmented Dickey-Fuller (ADF) test results for the time-series of the new users who interacted through quote tweet with content from Musk and Biden, respectively. \textbf{Right.} ARMA$(p,q)$ optimal parameters as results from exhaustive grid search with the Akaike Information Criterion (AIC).}
    \label{tab:ADFtest}
\end{table}

\begin{figure}[ht]
    \centering
\includegraphics[width=\textwidth]{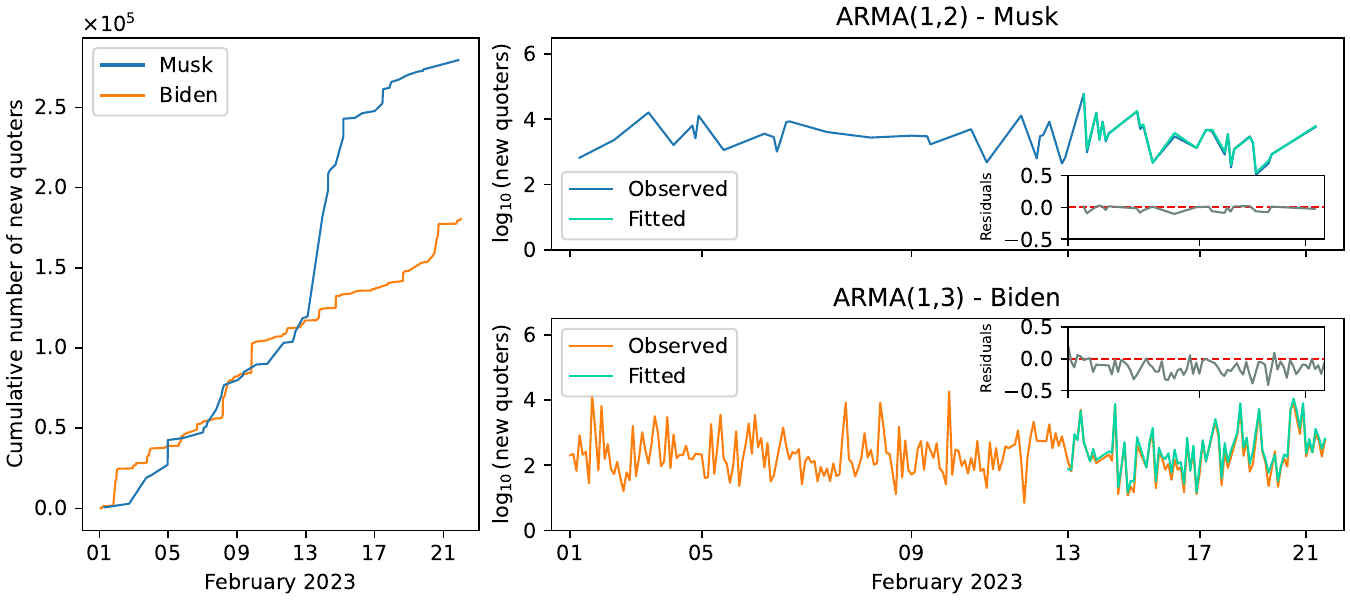}
    \caption{\textbf{Left.} Cumulative number of new users quoting content produced on Twitter by Elon Musk and Joe Biden, respectively. \textbf{Right.} Time-series of new quoters per tweet from Musk (top) and Biden (bottom), respectively. Main plots also show the prediction provided by the corresponding ARMA model (green lines). Inset plots show the trends of the residuals obtained by comparing real and predicted data (Observed - Fitted).}
    \label{fig:NewQuoters}
\end{figure}

The left panel of Fig.~\ref{fig:NewQuoters} shows that the cumulative number of new quoters of Biden follows a linear trend throughout the analysed period. While the number of Musk's new quoters practically overlaps with Biden's in the period preceding the algorithm changes announcement, it exhibits a sudden increase in the days immediately following and then returns to the previous trend. Nevertheless, for both accounts, the real observations from the test data period very rarely and by very little exceed the model predictions (Right panel of Fig.~\ref{fig:NewQuoters}).

On the Biden side, the model accuracy is consistent with the linear trend observed for the newly engaged users. 
Conversely, and coherently with what was observed in the previous section, the steep increase in the cumulative number of new quoters of Musk does not correspond to an increase in the probability of interaction with his content. 
In other words, many new users were involved, apparently only due to the increased visibility.
Analogous results concern the communities of retweeters (see SI). 

\subsection{External impact}
\begin{figure}[ht]
    \centering
\includegraphics[width=.45\textwidth]{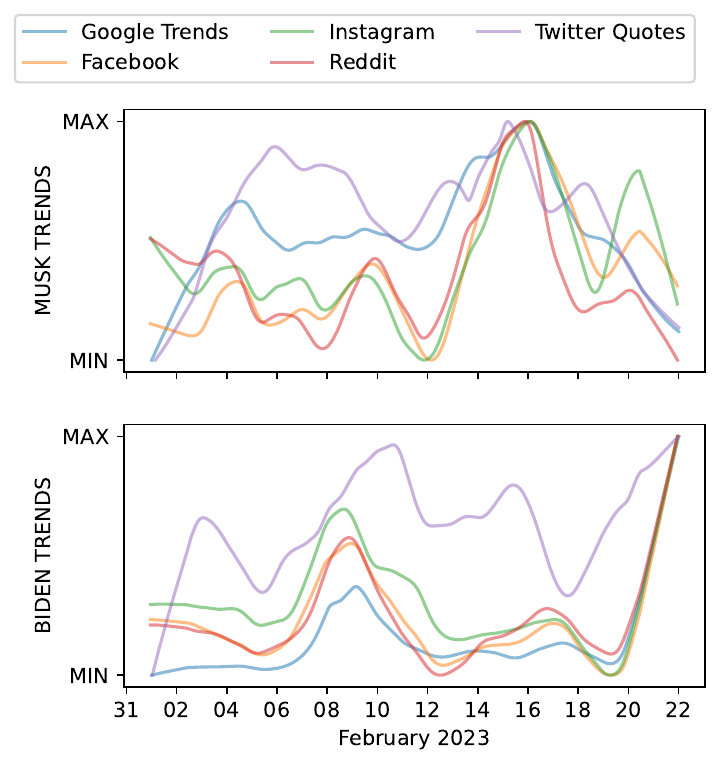}
\includegraphics[width=.45\textwidth]{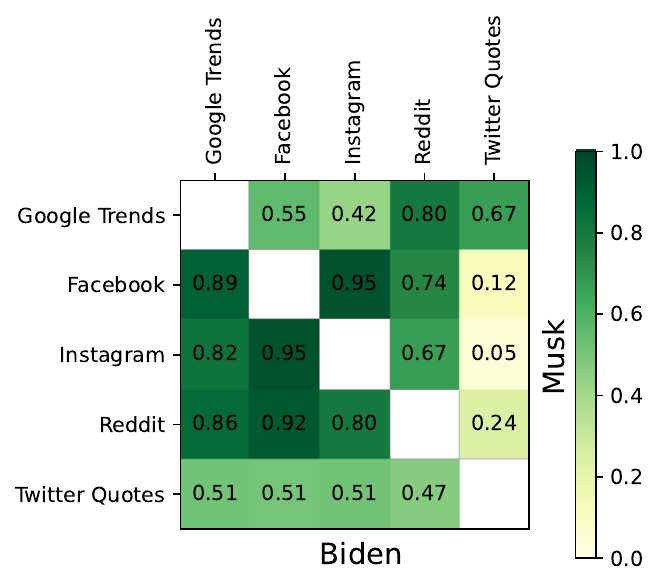}
\includegraphics[width=.7\textwidth]{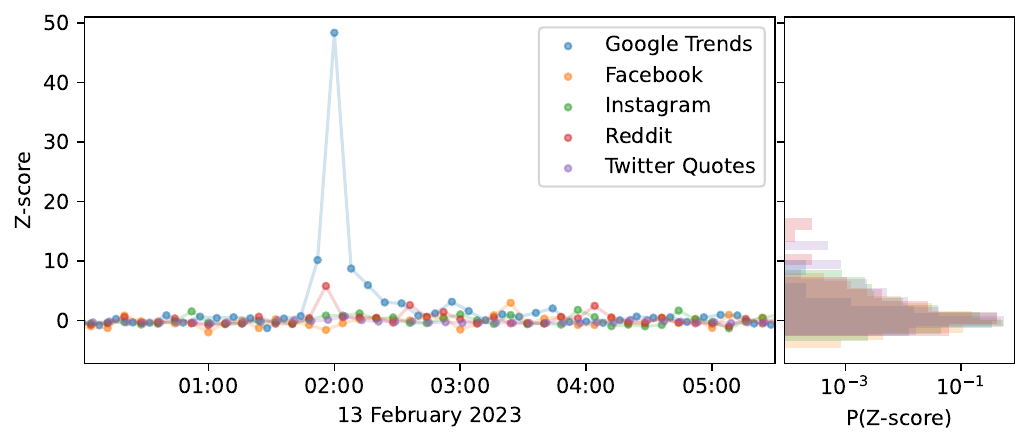}
    \caption{\textbf{Top left.} The deseasonalised time-series of the trends of the volume of contents concerning Musk and Biden, rescaled between the maximum and the minimum in the studied timeframe. \textbf{Top right.} The Spearman's correlation matrix of the time-series of trends. In the top right triangle, we reported the results for Musk correlations, while in the bottom left triangle, we reported the results for Biden. In all cases, $p-val < 0.05$. \textbf{Bottom.} The z-score of the residuals of deseasoning for the monitored social media time-series on the night of the 13th of February. The histogram for the whole period (without the outlier) can be observed in the right part.}
    \label{fig:extern}
\end{figure}

The impact inside the platform directly touched by the algorithm adjustments is, at least to some extent, expected and understandable. On the other hand, it is less obvious to study if and how that impact propagated to other platforms and, more in general, to the social dialogue.
For this purpose, we monitored Musk and Biden-related content on different platforms in the same timeframe: Facebook, Instagram, Reddit and Google Trends (see Methods for more details). 
The latter is not a social media like the others but it can be a powerful tool to assess the demand for information and the open questions in the social dialogue \cite{gravino2022supply}. 
We elaborated the deseasonalised time-series of the trends of the social dialogue involving the observed accounts of our research (see Methods).
To compare with what was happening on the platform, we also included the time-series of the volume of quote tweets obtained by the observed accounts inside the platform. We reported the results in the top left panels of Fig.~\ref{fig:extern}.
The curves exhibit a certain degree of coherence between all curves for both the observed accounts.
This coherence validates their adoption as proxies or projections of the public debate around the observed accounts.
In particular, we observe a common peak for Musk in the days when the algorithmic adjustments improved the visibility of his tweets (13-16 of February).
To give a more precise account of these similarities, we calculated the Spearman's correlation matrix between the time-series, and reported it in the top right part of Fig.~\ref{fig:extern}. 
Correlations are always strongly positive, except for the Twitter Quotes time-series with social media (Reddit, Instagram, and Facebook) for Musk. 
Musk's Twitter Quotes time-series show a strong correlation only with Google Trends, so we decided to deepen the analysis of this relation by looking at the residuals of the deseasoning. 
Residuals are mostly stable and time-independent, with one large deviation (z-score $\sim 50$) occurring for a few minutes during the night of the 13th of February. 
The detail of the residual z-scores for that night is reported in the bottom part of Fig.~\ref{fig:extern}, together with the z-score distribution on all the time-series. 
To complete the analysis, we assessed the influence of Musk's tweets directly by looking at the frequency variation in the social dialogue (i.e. in the posts on social media concerning him) of the words he used in his tweets. 
To estimate the influence, the words' occurrences (i.e. the number of appearance) in the days after the tweet was compared with the occurrences of the same words before the tweet (see Methods for details). 
We compared the period of improved visibility of his tweets (13-16 of February) with the previous period (5-12 February).
Results are reported in Table~\ref{tab:influence}, where it can be observed how a meaningful higher influence seems to emerge in the high visibility period, at least for some social media. We also report the variation of Occurrence (see Methods for details) to give an intuitive measure of the difference of Influence between the compared periods. The results show a strong increase in the Occurrence of tweets' words in all three social media posts, particularly Instagram and Facebook.

\begin{table}[ht]
    \centering
    \begin{tabular}{l|l|l|l}
        & Reddit & Instagram & Facebook \\
        \hline
        Kolmogorov-Smirnov statistics & $0.10$ & $0.26$ & $0.36$ \\
        Kolmogorov-Smirnov p-value& $0.548$ & $5.43e-4$ & $4.66e-7$ \\
        Occurrence Variation & $+10\%$ & $+70\%$ & $+62\%$ \\
    \end{tabular}
    \caption{The Kolmogorov-Smirnov comparison of the influence of Musk's tweet on the public discourse. The Occurrence Variation shows the percentual increase in the volume of appearance of the tweets' words in the public discourse in the following days. The high visibility period (13-16 February, 80 tweets) was compared with the previous period (5-12 February, 194 tweets). The $p-value$ below the threshold of $5\%$ for Instagram and Facebook show that the variation is significant.}
    \label{tab:influence}
\end{table}

To sum up, we observe a multifaceted external impact of the algorithmic adjustments. 
Comparing Musk and Biden, we observe different correlations between the public discourse on Twitter and other platforms. 
Twitter Quotes for Musk emerge as an outlier in the correlation metrics.
Also, we observe that Google Trends seem to be directly and strongly impacted by the algorithmic adjustments, at least in a punctual, direct way, and in an indirect way, given his strong correlation with Quotes, which, as we have seen before, increased for Musk during the algorithmic verification.
An impact can also be observed in the social dialogue, looking at the semantic influence of Musk's tweets on the social media content created by news outlets. 

\subsection{Community reaction}

\begin{figure}[ht]
    \centering
    \includegraphics[width=\textwidth]{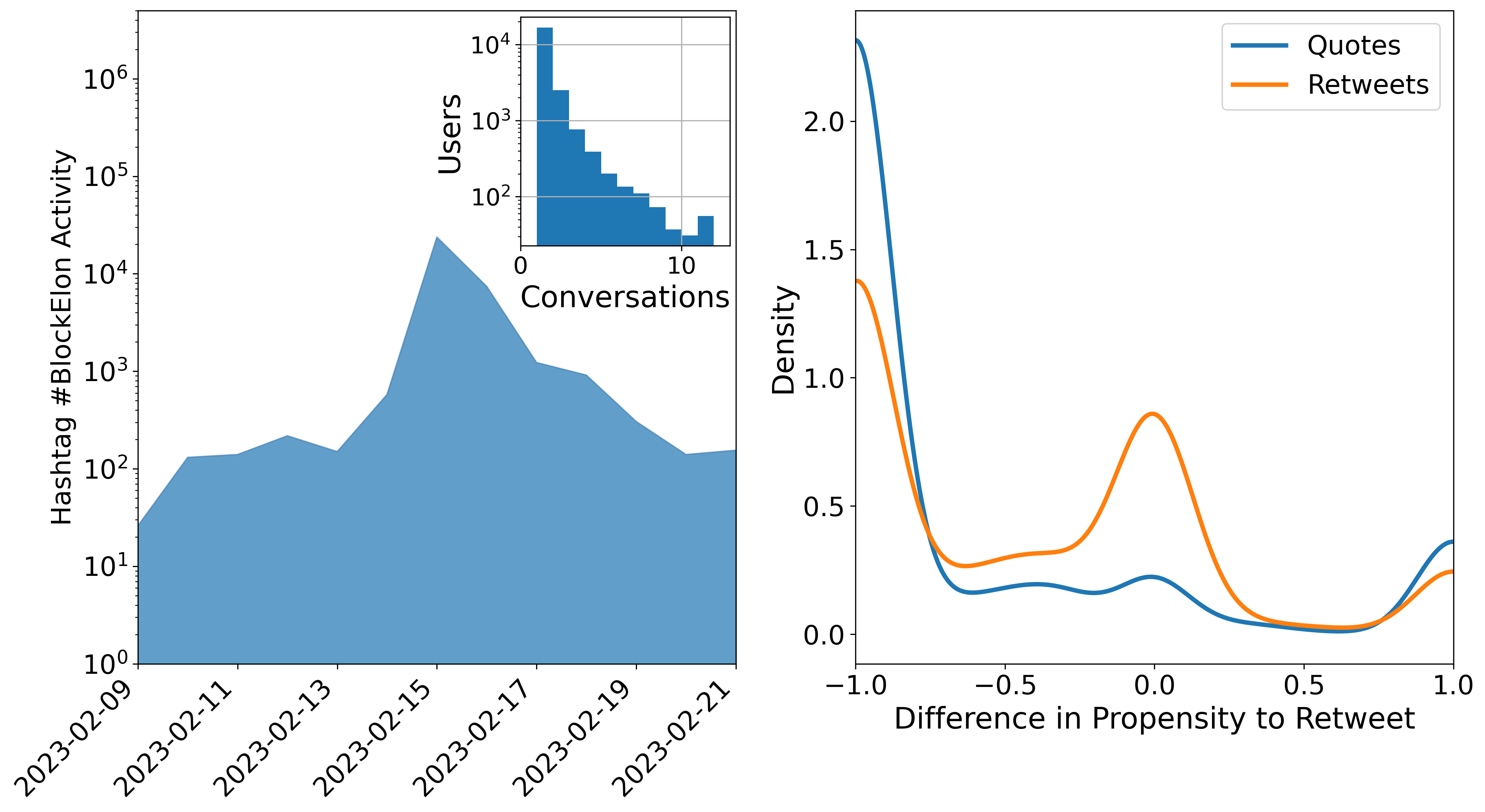}
    \caption{\textbf{Left.} The figure shows the number of tweets containing the hashtag \#BlockElon over the observed period of time. The inset displays the distribution of users against the number of different conversations in which they used the hashtag.
    \textbf{Right.} The figure shows the change in the propensity to \emph{quote} or \emph{retweet} Elon Musk after having used the hashtag \#BlockElon for the first time.}
    \label{fig:BESum}
\end{figure}

This part of the study investigated the impact of a social media campaign aimed at blacklisting Elon Musk on Twitter emerged as a spontaneous community reaction to the first critical reports of the algorithmic adjustments. To accomplish this, we focused on tweets from two distinct sources: (1) tweets posted by Elon Musk, (2) tweets containing the hashtag \#BlockElon.
We observe that the community using the hashtag existed before the period of peak visibility for Musk but that their activity increases by more than one order of magnitude after the algorithm changes (See Fig~\ref{fig:BESum}). 

To verify that participation in the community moved beyond the mere use of the hashtag, we measured the propensity to retweet and quote Elon Musk's tweets after adopting the \#BlockElon hashtag. 
This measurement will account for the coherence in the users' behaviour.
The propensity, defined as the variation in the probability of performing a certain action before and after the use of the hashtag, and normalized to the $[0,1]$ interval by dividing by the sum of probabilities (see Section~\ref{sec:propensity}), provides valuable insights into user behaviour. 
As Fig.~\ref{fig:BESum} illustrates, the propensity for users who have used the hashtag tends to be low, with global maxima for both retweets and quotes occurring at $-1$, indicating that many users do not engage in new retweets or quotes of Elon Musk's tweets after adopting the hashtag. 
Interestingly, the propensity distribution for quoting differs notably from that of retweeting, with the former exhibiting more extreme values than the latter. 
Additionally, a local peak in the propensity for retweeting around $0$ suggests that a segment of the community does not substantially alter their behaviour following hashtag adoption.

\begin{figure}[ht]
    \centering
\includegraphics[width=\textwidth]{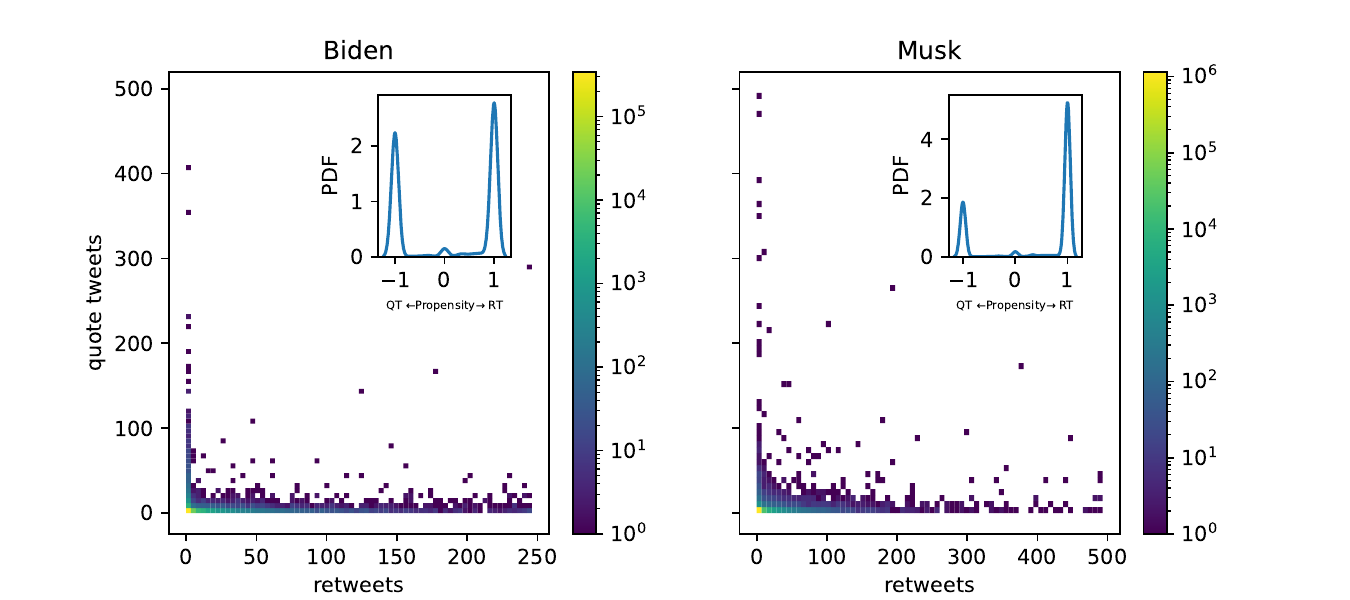}
    \caption{Relationship between number of retweets and number of quote tweets per user. The two variables appear to be strongly anti-correlated. This suggests that users interact with content produced on Twitter by Biden (Left) or Musk (Right) mainly through retweets or quote tweets and very rarely using both in a similar way. Insets show the bimodal distribution of the propensity score.}
    \label{fig:cor_quote_retweet}
\end{figure}

Moving beyond the \#BlockElon community, we conducted a similar analysis on the propensity to either retweet or quote Elon Musk's tweets compared to those of President Biden. 
Our findings reveal a distinctive pattern: when a user retweets one of the two accounts, it is unlikely that they will also quote the same account. 
The main plots of Fig.~\ref{fig:cor_quote_retweet} show the relationship between the number of retweets and the number of quote tweets per user for both accounts. 
The two variables appear to be strongly anticorrelated, as confirmed by Spearman's $\rho$ equal to $-0.67$ for Biden and $-0.60$ for Musk. 
The propensity, which ranges from -1 (indicating a preference for quoting) to 1 (indicating a preference for retweeting), with 0 denoting an equal probability of both actions, appears to be concentrated at the extremes, as shown in the insets of Fig.~\ref{fig:cor_quote_retweet}. 
This result suggests that users prefer retweeting or quoting rather than engaging in both actions simultaneously, at least regarding these two prominent Twitter accounts.

\section{Discussion}\label{discussion}

Recommender systems are gatekeepers of knowledge, but studying their impact on the public discourse is challenging due to the opaqueness of their implementation and evolution.
Thanks to the exceptionally transparent adjustments in the recommender algorithm of Twitter that occurred in February 2023, we could study the cross-platform impact of such changes on the public discourse, specifically on Twitter, Facebook, Instagram, Reddit, and Google Trends.
We focused on the visibility of two specific accounts since they were the main subjects of the journalistic reconstructions of the events: the account of Elon Musk, the platform owner, and the account of Joe Biden, the president of the United States, used as a benchmark.
Despite the lack of precise details on when the algorithm changes were made effective, thanks to the analysis of Twitter metrics, we could measure how the visibility of Musk's tweets increased substantially between the 13th and the 16th of February, 2023.
Such an increment was not observed for the benchmark.
During the same days, increased interactions (likes, retweets, etc.) were observed only for Musk, while the probability of interaction did not show any alteration. 
This rules out the possibility that increased visibility resulted from tweets being particularly engaging and, therefore, more promoted by the algorithm.
The same conclusion is reached when looking at the engaged communities.
We observed a steep increase in the cumulative number of new users engaged with content from Musk during the anomaly, while no changes concern the benchmark. Still, the number of newly engaged users per tweet in the days following the algorithm changes is accurately fitted by an ARMA model trained exclusively on data collected in the previous period.
This confirms that the probability of involvement remained stable and points to the enhanced visibility as the original cause of all observed anomalies.
To sum up, the announced adjustments strongly increased the visibility of content from Musk, at least for a few days, while apparently leaving other users' visibility unchanged.
This increase, in turn, augmented the engagement and the amount of new users engaged only for his tweets.

The consequences of this anomaly transcended the limits of Twitter. 
We analysed the social discourse about Musk and Biden on different platforms (Reddit, Google Trends, Facebook, and Instagram) and studied their link with the internal dynamics of Twitter through the time-series of quotes to the accounts' tweets.
All the time-series showed high coherence for both Musk and Biden. 
One important exception is Musk's Twitter Quotes time-series, which does not show any meaningful correlation with the social discourse on the social media platforms (Reddit, Facebook and Instagram) and a high correlation only with Google Trends. 
This difference in the correlations might be an effect of the changes to the algorithm. 
In other words, the normal relations between social dialogue instances on different social media might have been altered on Twitter by the increased visibility of the Musk account.

On the other hand, Google Trends shows a high correlation between Twitter Quotes for Musk and an extreme outlier, corresponding to a higher-than-usual volume of searches about Musk occurring during the high visibility period.
In particular, the extreme outlier in Google Trends cannot be explained as part of a normal increase of attention in the news cycle. 
It is an outlier of the trend that already considers that increase. 
Also, its time scale (only a few minutes) is inconsistent with the dynamics of a news cycle.
This impact can be explained as a reaction of the users that, exposed to the unusually high visibility of Musk tweets in their feeds (even if they were not following him), would ``google'' him and/or the bizarre omnipresence. 
Also, in the days of the anomaly, we observed a peak in the social debate about Musk on all platforms. 
This peak was probably due to his increased visibility since we observed, at least on Instagram and Facebook, a meaningfully stronger presence in the social debate of the words appearing in Musk's tweets.
This suggests that the increase in Musk's tweets' visibility due to the algorithmic adjustments managed to influence, at least for a few days, the social dialogue, causing users to ``google'' him and the media to talk about him.

Among the subtopics that emerged in the social dialogue about Musk were the algorithmic adjustments and the emergence of the \#BlockElon hashtag.
The latter was an apparently spontaneous community's reaction to some accusations\cite{platformer} and, probably, to the increased visibility itself and was part of a campaign to make the platform users block the account of Musk.
The hashtag became trending and peaked on the 15th of February, showing a delay compared to the peak of visibility of $1-2$ days, which can be considered as the reaction time of the community.
Users adopting the hashtag largely stayed coherent with their initial intent, refraining from quoting Musk's tweets after incorporating the \#BlockElon hashtag. 
This behaviour underscores a sense of collective purpose and commitment to the cause. 
The community's ability to stay on message suggests a level of organization and shared values, potentially contributing to their effectiveness in spreading their message.

Interestingly, the results are less apparent but intriguing when considering the impact on the propensity to retweet. 
It appears that quoting Musk's tweets tends to be more common among users who hold a more critical stance, while retweets may involve a broader range of followers, including those who might not necessarily endorse the \#BlockElon movement but still engage with the content. 
This nuanced differentiation in engagement tactics within the community provides valuable insights into how social media users express their sentiments and actively participate in shaping the discourse around influential figures like Elon Musk. 
Notably, a striking observation emerges when extending the analysis beyond the \#BlockElon community. 
Users who quote the accounts of President Biden or Musk almost never retweet the same accounts. 
This divergence in behaviour highlights the stark contrast between quoting and retweeting practices, even when not specifically focused on the \#BlockElon movement. 
This is consistent with the original intent behind introducing the quote in April 2015~\cite{garimella2016quote}. 
Quotes were introduced to Twitter as a response to the community's request for a tool to express more critical positions with respect to a tweet. 
Our analysis shows that this behaviour is still consistent with the original intent of the tool. In future work, the different intrinsic values of these reactions, or similar reactions on different platforms, can open new directions in studying how users and communities interact (positively or negatively) on social networks.

\section{Conclusion}
In digital societies, algorithms can impact citizens and media behaviours, and control over them is often opaque, posing serious issues for the health of the social dialogue and, therefore, democracy.
The exceptional situation of the announced adjustments to the Twitter recommender algorithm in February 2023 allowed us to demonstrate that the increased visibility of an individual account can directly impact citizens and media, driving the social dialogue and triggering criticisms and protests.
In most other platforms, algorithms and their development are often opaque.
Possible attempts to manipulate the public discourse would not be announced and would be harder to notice.
Our work also gives insights into anomaly detection through cross-platform analysis, through the correlation analysis, or the outliers analysis. 

Still, our work suffers from several limitations.
Some will be addressed in future works, while others seem more challenging.
A larger set of accounts could be studied to define the benchmark better, and the analysis can be expanded in time.
Also, the impact of individual tweets on the social dialogue can be evaluated by adopting more sophisticated NLP tools.
Generalising the results poses more serious problems. 
On the one hand, the traditional lack of transparency in the management of recommender systems makes the occurrence of a similar event unlikely.
On the other hand, recent Twitter changes in data access policies for research make it harder to perform similar studies.
Transparency of algorithms' management and data availability for research are the two pillars we leveraged in our work, and we believe they will be vital for any future work on the health of the social dialogue and, therefore, of democracy.

\section{Methods}\label{methods}

\subsection{Data collection}
We exploited the official Twitter API to collect all content (tweets, quoted tweets, retweets, and replies) posted on the platform by Elon Musk and Joe Biden from 1 to 21 February 2023 and corresponding retweets and quoted tweets as of 21 March.

To investigate the impact of the BlockElon campaign, we also gathered all content, including one of the following hashtags: \textit{\#blockelon, \#blockmusk, \#blockelonmusk, \#muteelon, \#mutemusk, \#muteelonmusk}.

To study how Musk's statements have influenced the public discourse on other platforms, we further exploited the CrowdTangle tool to collect all publicly available content published on Facebook, Instagram, and Reddit in the same timespan, whose textual components include one of the keywords listed in Table \ref{tab:other_platforms_kw}.

\begin{table}[ht]
    \centering
    \begin{tabular}{l|l}
        Biden-related & Musk-related \\
        \hline
        \textit{biden, ``joe biden'', ``president biden'',} & \textit{``elon musk'', musk, @elonmusk,}\\
        \textit{\#joebiden, \#potus, @potus, ``president} & \textit{\#elonmusk, ``ceo of twitter'',}\\
        \textit{of the us'', ``president of the u.s.'',} & \textit{``twitter ceo'', ``c.e.o. of twitter'',}\\
        \textit{``us president'', ``u.s. president'',} & \textit{``twitter c.e.o.''}\\
        \textit{``president of the united states''} & 
    \end{tabular}
    \caption{List of keywords used to filter Biden-related and Musk-related content on Facebook, Instagram, and Reddit.}
    \label{tab:other_platforms_kw}
\end{table}
To avoid including erroneous content with `musk' referring to topics other than Musk person, content containing only the word `musk' was further limited to those containing: \textit{``mr musk'', ``musk foundation'', musk AND elon, musk AND zuckerberg, musk AND starlink, musk AND spacex, musk AND tesla, musk AND murdoch, musk AND twitter, musk AND chatgpt, musk AND starship, musk AND tweet, musk AND bezos, musk AND ``bill gates''}.

Tables~\ref{tab:twitter_breakdown} and \ref{tab:other_breakdown} show a breakdown of Twitter and other-platform datasets, respectively.
\begin{table}[ht]
\footnotesize
    \centering
    \begin{tabular}{c||ccccc||ccc}
         & \multicolumn{5}{c||}{Production} & \multicolumn{3}{c}{Consumption}\\
         \hline
         & User & TW & QT & RE & RT & User & QT & RT\\
Biden & 1 & 224 & 14 & 12 & 0 & 384,499 & 323,345 & 771,801\\
Musk & 1 & 54 & 26 & 419 & 27 & 1,178,235 & 495,896 & 1,979,080\\
\#BlockElon & 21,202 & 8,401 & 0 & 4,302 & 22,486 & --- & --- & ---
    \end{tabular}
    \caption{Breakdown of data from Twitter. TW=tweet, QT=quote tweet, RE=reply, RT=retweet.}
    \label{tab:twitter_breakdown}
\end{table}

\begin{table}[ht]
\small
    \centering
    \begin{tabular}{c||c||ccccc}
         & User / & Link & Photo & Status & Video & Total\\
         & Subreddit &  &  &  &  & \\
         \hline
Facebook & 51,428 & 138,988 & 50,430 & 6,130 & 29,425 & 224,973\\
Instagram & 10,786 & --- & 24,442 & --- & 735 & 25,177\\
Reddit & 1,276 & 9,885 & 656 & 1,897 & 599 & 13,037
    \end{tabular}
    \caption{Breakdown of data from other social media. W.r.t. to Facebook, Video also includes Live Video,  Live Video Complete, Live Video Scheduled, Native Video, and YouTube.
W.r.t. to Instagram, Photo also includes Album.
W.r.t. to Reddit, Video also includes YouTube.}
    \label{tab:other_breakdown}
\end{table}

\subsection{Deseasoning: trends and residuals}
The time-series of the content volume concerning the observed accounts on the different social networks was analysed after a deseasoning elaboration to study separately trends and residuals.
The deseasoning procedure has been simply implemented leveraging the python library \textit{statsmodel.tsa.seasonal} and the function \textit{STL}, which is built on and improves a previous research work~\cite{cleveland1990stl}. 
The function gives back the elaboration of the deseasoned trend time-series and of the residuals time-series.
Trends have been treated by rescaling them between the maximum and the minimum in the observed interval to make them comparable with Google Trends data at least to some extent.
These data, in fact, by nature, is lacking an absolute unit of measure.
For the same reason, the correlation analysis has been performed by elaborating Spearman's coefficient, which does not compare the absolute values but the ranks. 

\subsection{Modelling the evolution of the user communities involved}

The evolution of the communities of users who interacted with content published by Elon Musk and Joe Biden, respectively, is assessed employing Autoregressive Moving Average (ARMA) models~\cite{durbin2012time}. The proposed models incorporate two exogenous variables aiming at balancing natural biases that can affect the estimates and then need to be considered: the reach of a tweet and its timestamp.
Denoted with $t_i$ the timestamp of tweet $i$, we relied on the total number of quote tweets of $i$ as a proxy for the former and the opposite of the cumulative count of quote tweets at time $t_i$ as a proxy for the latter.
Indeed, a higher number of quote tweets increases the account's visibility, hence positively impacting the likelihood of attracting new quoters who may not have previously engaged with the account’s content. For instance, from our viewpoint, the quoters of the first tweet are all new quoters, although they have almost certainly already quoted the reference account before our observation period.
Conversely, as time progresses, the active audience of the account tends to be fully engaged, hence negatively affecting the number of new quoters.
Analogous considerations hold considering retweet as reference interaction, rather than quote tweet.
Hence, the proposed ARMA model is defined as:
\begin{equation}  Y_t=\beta_1Y_{1,t}+\beta_2Y_{2,t}+\displaystyle\sum_{i=1}^p\phi_iY_{t-i}+\displaystyle\sum_{i=1}^q\theta_i\epsilon_{t-i}+\sigma^2
\end{equation}
where
\begin{itemize}
    \item $Y_t$ is the observed time-series at time $t$;
    \item $Y_{i,t}$ and $\beta_i$ are the exogenous variables at time $t$ and the corresponding coefficients, respectively;
    \item $\phi_1,\dots,\phi_p$ are autoregressive coefficients;
    \item $\theta_1,\dots,\theta_p$ are moving average coefficients;
    \item $\sigma^2$ is the variance of the residual values.
\end{itemize}
The stationarity of the time-series is assessed through augmented Dickey–Fuller test (ADF)~\cite{said1984}.
To enhance the suitability of the analysis, data is regularized via $x\mapsto \log_{10}(x+1)$ transformation before fitting.
Further, the parameters of models are estimated using the Akaike Information Criterion (AIC) for optimal fit~\cite{akaike1974}.

We train the models on data before the announcement of the algorithm changes (13 February 2023), and then we compare the 
subsequent real data with the model's predictions in order to test whether or not a clear change emerges from the dynamics of new quoters (or retweeters).
The modeling process has been conducted using the python library \textit{statsmodels.tsa.statespace.sarimax} and the function \textit{SARIMAX}.

\subsection{Influence estimation}

To estimate the punctual impact of tweets on social media public discourse, we defined Influence $I$ of tweet $x$ as:

\begin{equation}
I(x) = log_{10}(O_{post}(x) + 1) - log_{10}(O_{pre}(x)+1)
\end{equation}

where $O$ is the Occurrences, which is calculated as follows. 
Given the set of words ${w}$ appearing in the tweet $x$ (filtered to exclude stopwords and less than 3 characters words), we count the number of the appearances of those words in social media posts in the 4 days preceding (for $O_{pre}$) and following (for $O_{post}$) the publication of $x$. 
We followed this procedure for the three social media included in our study (Reddit, Instagram, and Facebook). We reported in Tab.~\ref{tab:influence} the results of the Kolmogorov-Smirnov test for the Influence of the tweets in the high visibility period, and in the rest of the monitored period (excluding the first and the last 4 days). We also report the percentual variation of Occurrences to make the results more understandable.

\subsection{\#BlockElon activity and propensity}
\label{sec:propensity}


We selected this hashtag as it was widely used by a group of users advocating for others to blacklist Elon Musk on Twitter. We chose to analyze the behaviour of users who used this hashtag before and after using it to determine whether or not it impacted their social media activity.
To this end, we define the propensity score as the variation in the probability of performing a given action (retweet and quote for this analysis) over the sum of the probability before and after using the hashtag.
\begin{equation}
    \Delta P = \frac{p_{\text{after}}(\text{action}/\text{Hashtag Use})-p_{\text{before}}(\text{action}/\text{Hashtag Use})}{p_{\text{after}}(\text{action}/\text{Hashtag Use})+p_{\text{before}}(\text{action}/\text{Hashtag Use})}
\end{equation}
Since users often use the hashtag multiple times, we take the time of use as the first detected hashtag use. We test for differences in the distribution of propensity by varying this prescription and find no significant changes.

\backmatter
\section*{Acknowledgments}
We thank Peter Hanappe and Vittorio Loreto for their comments and suggestions that helped us improve the manuscript. This work has been supported by the Horizon Europe VALAWAI project (grant agreement number 101070930).

\section*{Data Availability Statements}\label{data}
All relevant data used in this study not protected by copyright are available upon reasonable request from the corresponding author.

\section*{Code Availability Statements}\label{code}
All relevant codes used in this study are available upon reasonable request from the corresponding author.

\section*{Competing interests}
The authors declare no competing interests.

\section*{Authors' contributions}
P.G. conceived and supervised the work. E.B. gathered and curated the data from social media. P.G. gathered and curated the data from Google Trends. All the authors analyzed the data. All the authors drafted, revised and wrote the manuscript. All authors participated in the discussions and agreed with the contents of this work.

\pagebreak
\bibliography{sn-bibliography}

\end{document}


\title{Cross-platform impact of social media algorithmic adjustments on public discourse}


\author*[1,3]{\fnm{Pietro} \sur{Gravino}}\email{pietro.gravino@sony.com}

\author[2,3]{\fnm{Ruggiero D.} \sur{Lo Sardo}}\email{ruggiero.losardo@sony.com}

\author[2,3]{\fnm{Emanuele} \sur{Brugnoli}}\email{emanuele.brugnoli@sony.com}


\affil*[1]{\orgname{Sony CSL - Paris}, \country{France}}

\affil[2]{\orgname{Sony CSL - Rome, Joint Initiative CREF-SONY, Centro Ricerche Enrico Fermi}, \city{Rome}, \country{Italy}}

\affil[3]{\orgname{Centro Ricerche Enrico Fermi}, \city{Rome}, \country{Italy}}


\section*{Supplementary information}


\subsection*{Like, retweet, quote counts and probabilities}
\begin{figure}[ht!]
    \centering
\includegraphics[width=.49\textwidth]{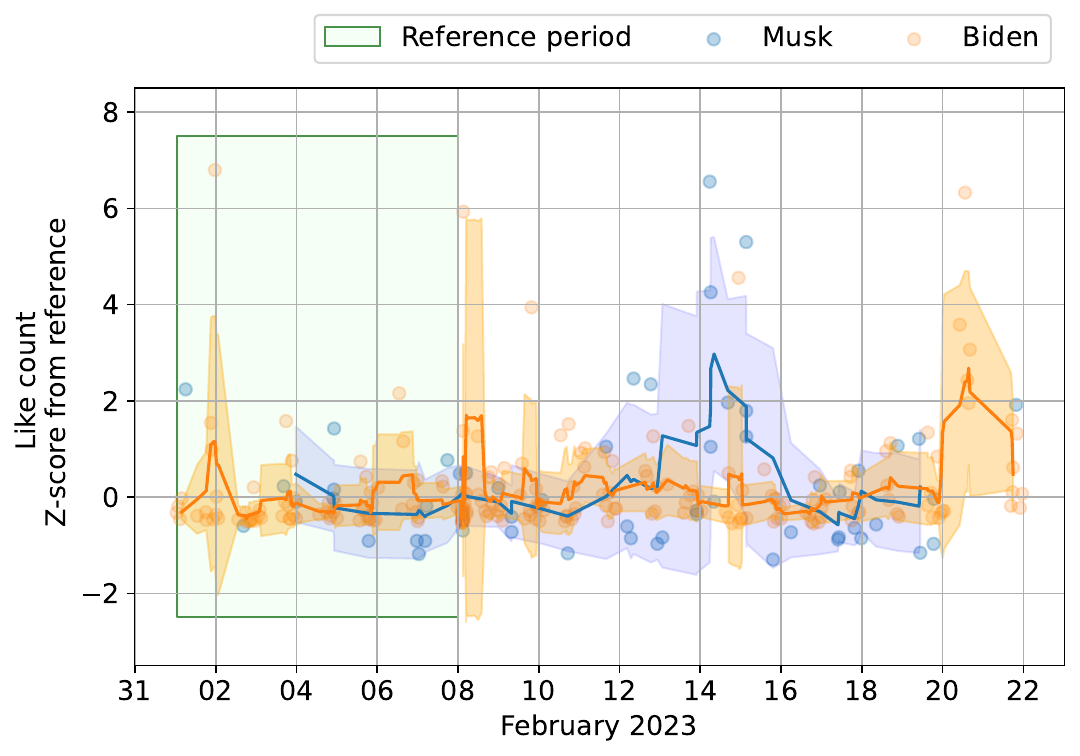}
\includegraphics[width=.49\textwidth]{fig/like_prob.pdf}
\includegraphics[width=.49\textwidth]{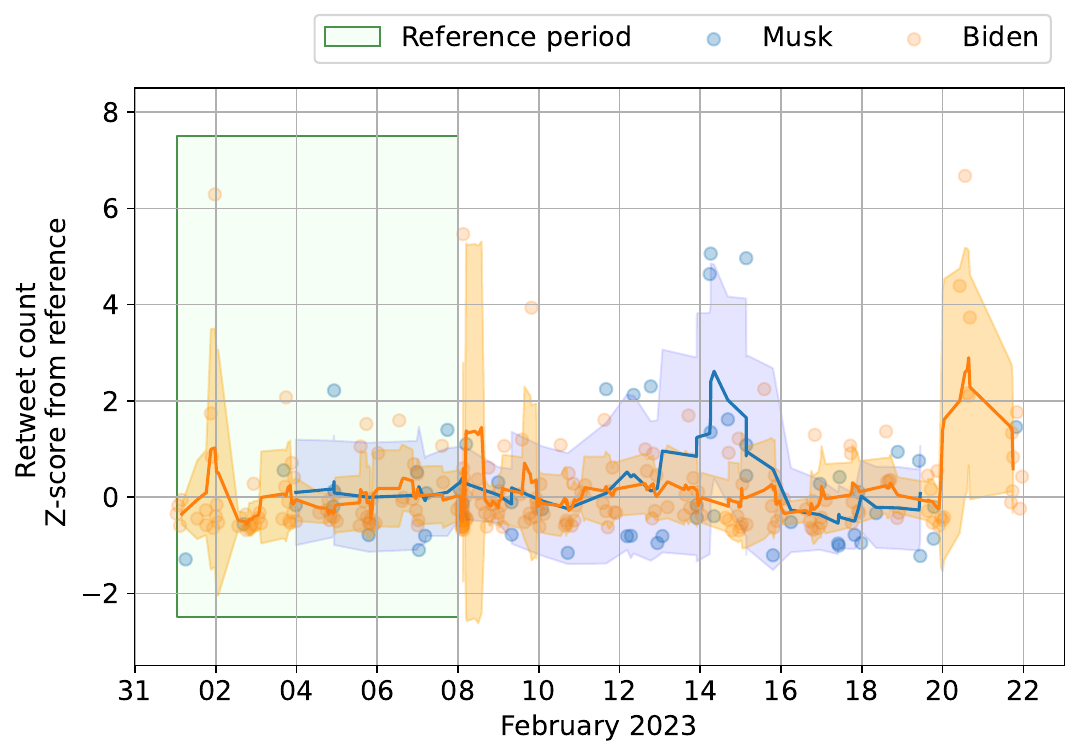}
\includegraphics[width=.49\textwidth]{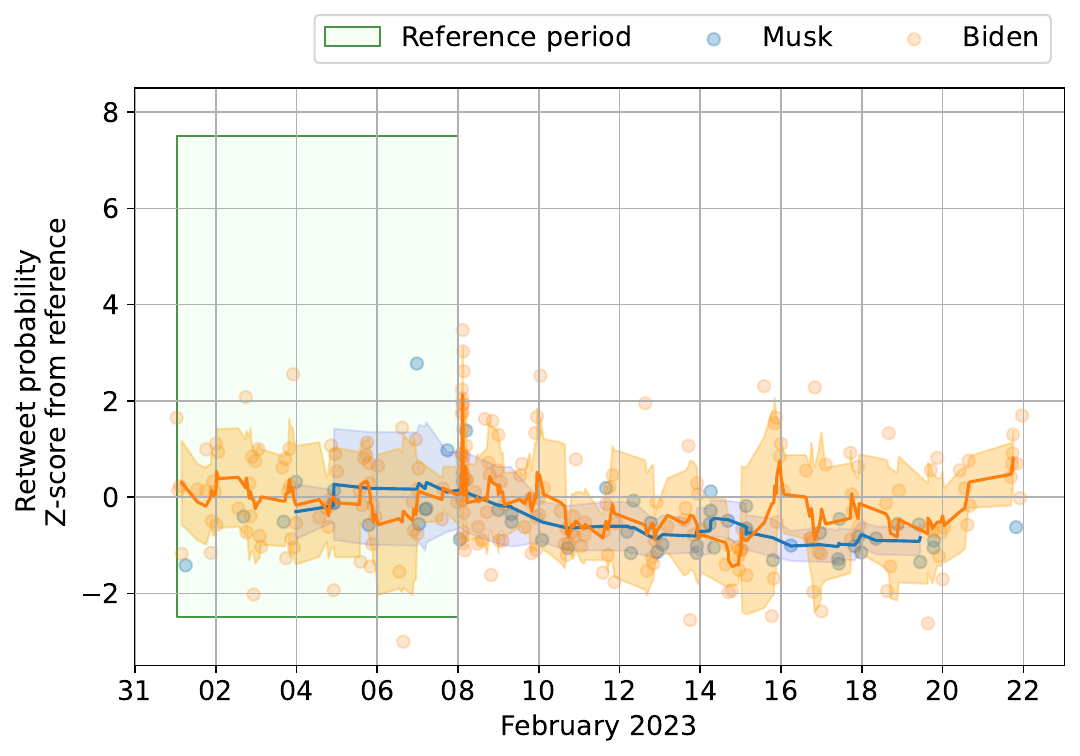}
\includegraphics[width=.49\textwidth]{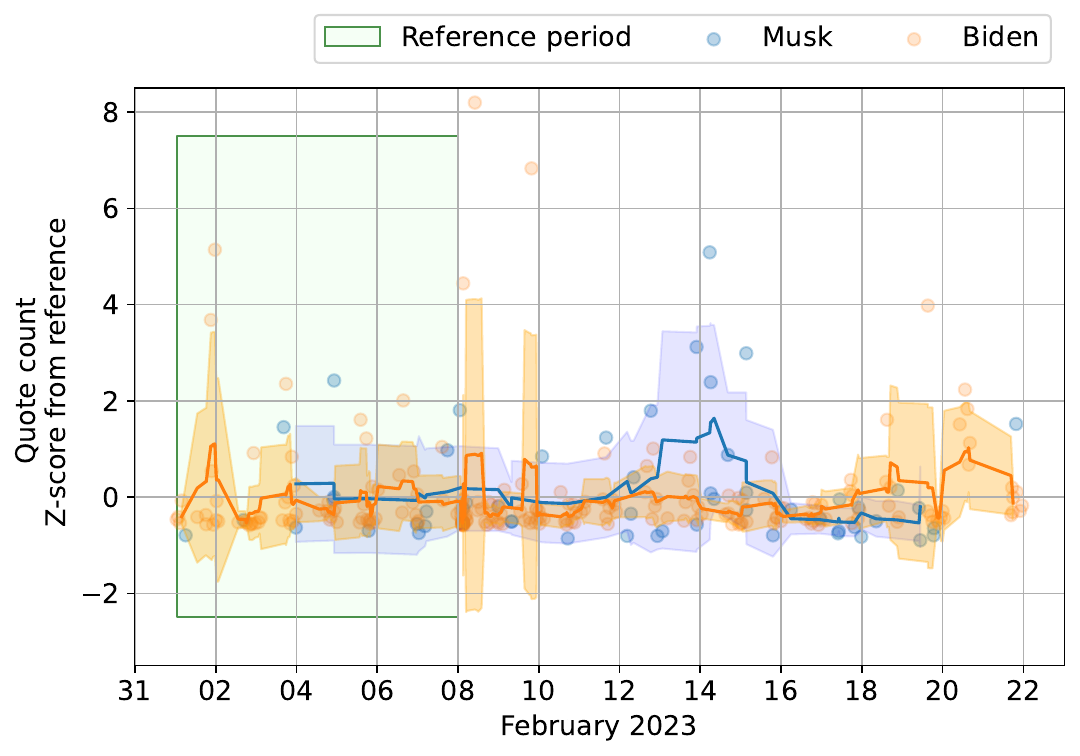}
\includegraphics[width=.49\textwidth]{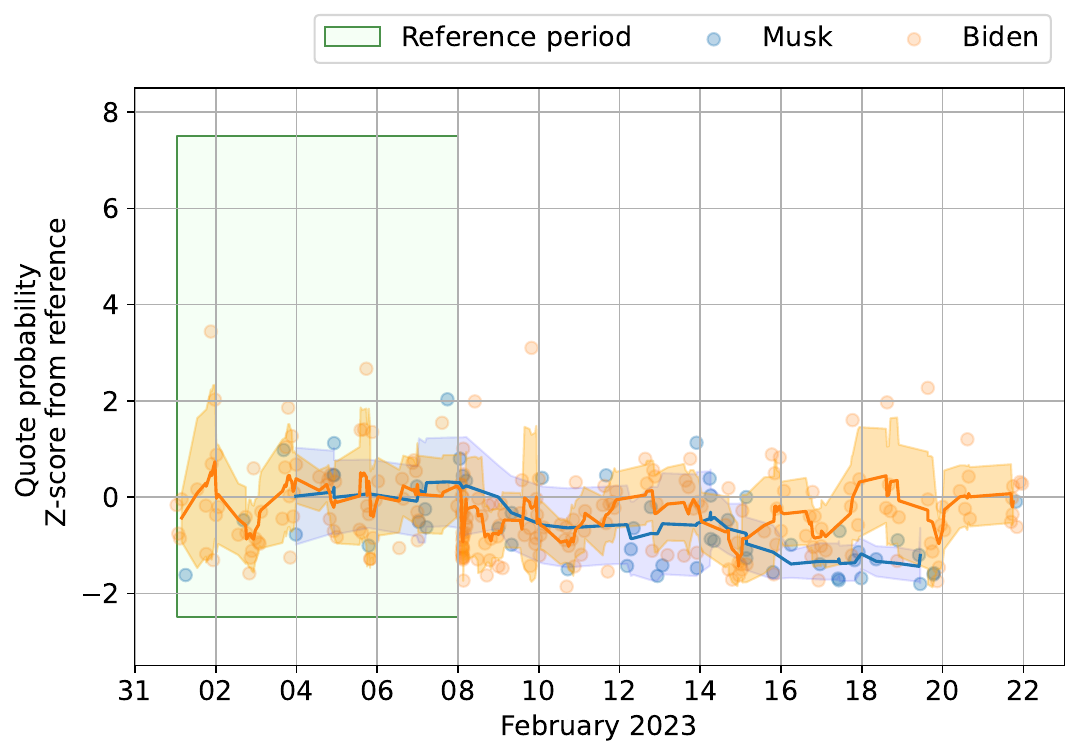}
    \caption{\textbf{Left column.} The z-score of the like, retweet, and quote count for tweets of Elon Musk and Joe Biden (benchmark). \textbf{Right column.} The z-score of the like, retweet, and quote probability for Elon Musk and Joe Biden (benchmark). \textbf{Both.} The line represents the running average, and the shade represents the standard deviation (window = 7 tweets). The reference period includes the first seven days of February 2023.}
    \label{fig:countnprob}
\end{figure}
\newpage

\subsection*{Internal impact on the communities of retweeters}
\begin{table}[ht!]
   \centering
   \begin{tabular}{c|cc|rr}
       & \multicolumn{2}{c|}{ADF} & \multicolumn{2}{c}{ARMA opt. pars}\\
       & Statistic & p-value & $(p,q)$ & AIC\\
       \hline
       Musk & $-6.40$ & $2.02e-08$ & $(3,2)$ & $-55.43$\\
       Biden & $-14.50$ & $5.98e-27$ & $(1,1)$ & $18.45$\\
   \end{tabular}
   \caption{\textbf{Left.} Augmented Dickey-Fuller (ADF) test results for the time-series of the new users who interacted through retweet with content from Musk and Biden, respectively. \textbf{Right.} ARMA$(p,q)$ optimal parameters as results from exhaustive grid search with the Akaike Information Criterion (AIC).}
   \label{tab:ADFtestRT}
\end{table}

\begin{figure}[ht!]
   \centering
\includegraphics[width=\textwidth]{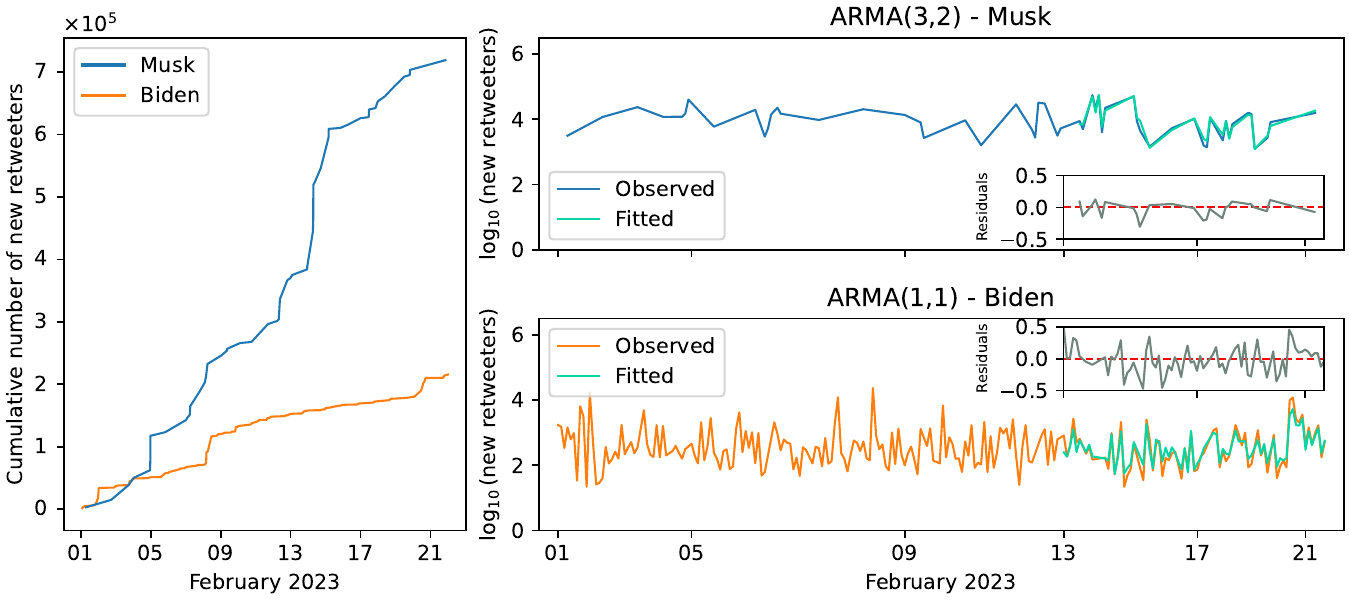}
   \caption{\textbf{Left.} Cumulative number of new users retweeting content produced on Twitter by Elon Musk and Joe Biden, respectively. \textbf{Right.} Time-series of new retweeters per tweet from Musk (top) and Biden (bottom), respectively. Main plots also show the prediction provided by the corresponding ARMA model (green lines). Inset plots show the trends of the residuals obtained by comparing real and predicted data (Observed - Fitted).}
   \label{fig:NewRetweeters}
\end{figure}
\newpage

\subsection*{Detailed results of the ARMA models}
\begin{table}[ht!]
   \centering
   \begin{tabular}{c|rr|rr}
   & \multicolumn{2}{c|}{quote tweets} & \multicolumn{2}{c}{retweets}\\
   & Biden & Musk & Biden & Musk\\
   \hline
   min & $-0.414$ & $-0.103$ & $-0.469$ & $-0.305$\\
   first quantile & $-0.221$ & $-0.061$ & $-0.140$ & $-0.100$\\
   median & $-0.134$ & $-0.009$ & $-0.026$ & $-0.004$\\
   mean & $-0.134$ & $-0.024$ & $-0.015$ & $-0.028$\\
   third quantile & $-0.048$ & $0.012$ & $0.109$ & $0.052$\\
   max & $0.223$ & $0.028$ & $0.516$ & $0.124$
   \end{tabular}
   \caption{Summary of relevant statistics on the residuals obtained comparing real data with the synthetic ones from the corresponding ARMA model.}
   \label{tab:residual_comparison}
\end{table}

\begin{table}[ht!]
    \centering
    \begin{tabular}{lclc}
    \toprule
    \textbf{Dep. Variable:}          &        y         & \textbf{  No. Observations:  } &    152      \\
    \textbf{Model:}                  & ARMA(1, 3) & \textbf{  Log Likelihood     } &  109.637    \\
    \textbf{Account:}                   & Biden & \textbf{  AIC                } &  -205.274   \\
    \textbf{Interaction:}                   &     Quote tweet     & \textbf{  BIC                } &  -184.106   \\
    \textbf{Covariance Type:}                 &        opg         & \textbf{  HQIC               } &  -196.675   \\
    \bottomrule
    \end{tabular}
    \begin{tabular}{lcccccc}
                & \textbf{coef} & \textbf{std err} & \textbf{z} & \textbf{P$> |$z$|$} & \textbf{[0.025} & \textbf{0.975]}  \\
    \midrule
    $\beta_1$     &       0.1488  &        0.014     &    10.328  &         0.000        &        0.121    &        0.177     \\
    $\beta_2$     &       1.1577  &        0.027     &    43.226  &         0.000        &        1.105    &        1.210     \\
    $\phi_1$  &       0.8041  &        0.187     &     4.289  &         0.000        &        0.437    &        1.172     \\
    $\theta_1$  &      -0.5434  &        0.207     &    -2.628  &         0.009        &       -0.949    &       -0.138     \\
    $\theta_2$  &      -0.0386  &        0.102     &    -0.379  &         0.705        &       -0.239    &        0.161     \\
    $\theta_3$  &       0.0434  &        0.114     &     0.382  &         0.703        &       -0.180    &        0.266     \\
    $\sigma^2$ &       0.0147  &        0.002     &     8.062  &         0.000        &        0.011    &        0.018     \\
    \bottomrule
    \end{tabular}
    \begin{tabular}{lclc}
    \textbf{Ljung-Box ($\theta_1$) (Q):}     & 0.00 & \textbf{  Jarque-Bera (JB):  } & 11.69  \\
    \textbf{Prob(Q):}                & 0.99 & \textbf{  Prob(JB):          } &  0.00  \\
    \textbf{Heteroskedasticity (H):} & 3.25 & \textbf{  Skew:              } & -0.50  \\
    \textbf{Prob(H) (two-sided):}    & 0.00 & \textbf{  Kurtosis:          } &  3.91  \\
    \bottomrule
    \end{tabular}
    \caption{Results of the ARMA model describing the dynamics of new quoters of Biden.}
\end{table}

\begin{table}[ht!]
    \begin{tabular}{lclc}
    \toprule
    \textbf{Dep. Variable:}          &        y         & \textbf{  No. Observations:  } &    152      \\
    \textbf{Model:}                  & ARMA(1, 1) & \textbf{  Log Likelihood     } &   -4.226    \\
    \textbf{Account:}                   & Biden & \textbf{  AIC                } &   18.452    \\
    \textbf{Interaction:}                   &     Retweet     & \textbf{  BIC                } &   33.571    \\
    \textbf{Covariance Type:}                 &        opg         & \textbf{  HQIC               } &   24.594    \\
    \bottomrule
    \end{tabular}
    \begin{tabular}{lcccccc}
                & \textbf{coef} & \textbf{std err} & \textbf{z} & \textbf{P$> |$z$|$} & \textbf{[0.025} & \textbf{0.975]}  \\
\midrule
$\beta_1$     &       0.3124  &        0.025     &    12.531  &         0.000        &        0.264    &        0.361     \\
$\beta_2$     &       1.2644  &        0.036     &    35.298  &         0.000        &        1.194    &        1.335     \\
$\phi_1$  &       0.9238  &        0.081     &    11.399  &         0.000        &        0.765    &        1.083     \\
$\theta_1$  &      -0.7530  &        0.132     &    -5.701  &         0.000        &       -1.012    &       -0.494     \\
$\sigma^2$ &       0.0641  &        0.009     &     7.493  &         0.000        &        0.047    &        0.081     \\
\bottomrule
\end{tabular}
\begin{tabular}{lclc}
\textbf{Ljung-Box ($\theta_1$) (Q):}     & 0.09 & \textbf{  Jarque-Bera (JB):  } & 10.37  \\
\textbf{Prob(Q):}                & 0.77 & \textbf{  Prob(JB):          } &  0.01  \\
\textbf{Heteroskedasticity (H):} & 0.92 & \textbf{  Skew:              } &  0.60  \\
\textbf{Prob(H) (two-sided):}    & 0.76 & \textbf{  Kurtosis:          } &  3.45  \\
\bottomrule
\end{tabular}
\caption{Results of the ARMA model describing the dynamics of new retweeters of Biden.}
\end{table}

\begin{table}[ht!]
\begin{tabular}{lclc}
\toprule
\textbf{Dep. Variable:}          &        y         & \textbf{  No. Observations:  } &     28      \\
\textbf{Model:}                  & ARMA(1, 2) & \textbf{  Log Likelihood     } &   57.960    \\
\textbf{Account:}                   & Musk & \textbf{  AIC                } &  -103.919   \\
\textbf{Interaction:}                   &     Quote tweet     & \textbf{  BIC                } &  -95.926    \\
\textbf{Sample:}                 &        0         & \textbf{  HQIC               } &  -101.476   \\
\bottomrule
\end{tabular}
\begin{tabular}{lcccccc}
                & \textbf{coef} & \textbf{std err} & \textbf{z} & \textbf{P$> |$z$|$} & \textbf{[0.025} & \textbf{0.975]}  \\
\midrule
$\beta_1$     &       0.0745  &        0.009     &     7.962  &         0.000        &        0.056    &        0.093     \\
$\beta_2$     &       1.0728  &        0.013     &    82.095  &         0.000        &        1.047    &        1.098     \\
$\phi_1$  &       0.7596  &        0.457     &     1.663  &         0.096        &       -0.136    &        1.655     \\
$\theta_1$  &      -0.6550  &        0.540     &    -1.212  &         0.225        &       -1.714    &        0.404     \\
$\theta_2$  &      -0.2092  &        0.216     &    -0.969  &         0.332        &       -0.632    &        0.214     \\
$\sigma^2$ &       0.0009  &        0.000     &     2.272  &         0.023        &        0.000    &        0.002     \\
\bottomrule
\end{tabular}
\begin{tabular}{lclc}
\textbf{Ljung-Box ($\phi_1$) (Q):}     & 0.18 & \textbf{  Jarque-Bera (JB):  } &  1.64  \\
\textbf{Prob(Q):}                & 0.67 & \textbf{  Prob(JB):          } &  0.44  \\
\textbf{Heteroskedasticity (H):} & 1.13 & \textbf{  Skew:              } & -0.06  \\
\textbf{Prob(H) (two-sided):}    & 0.86 & \textbf{  Kurtosis:          } &  1.82  \\
\bottomrule
\end{tabular}
\caption{Results of the ARMA model describing the dynamics of new quoters of Musk.}
\end{table}

\begin{table}[ht!]
\begin{tabular}{lclc}
\toprule
\textbf{Dep. Variable:}          &        y         & \textbf{  No. Observations:  } &     28      \\
\textbf{Model:}                  & ARMA(3, 2) & \textbf{  Log Likelihood     } &   35.716    \\
\textbf{Account:}                   & Musk & \textbf{  AIC                } &  -55.433    \\
\textbf{Interaction:}                   &     Retweet     & \textbf{  BIC                } &  -44.775    \\
\textbf{Sample:}                 &        0         & \textbf{  HQIC               } &  -52.175    \\
\bottomrule
\end{tabular}
\begin{tabular}{lcccccc}
                & \textbf{coef} & \textbf{std err} & \textbf{z} & \textbf{P$> |$z$|$} & \textbf{[0.025} & \textbf{0.975]}  \\
\midrule
$\beta_1$     &       0.2276  &        0.061     &     3.739  &         0.000        &        0.108    &        0.347     \\
$\beta_2$     &       1.2313  &        0.076     &    16.149  &         0.000        &        1.082    &        1.381     \\
$\phi_1$  &      -0.8334  &        0.805     &    -1.035  &         0.301        &       -2.412    &        0.745     \\
$\phi_2$  &      -0.2549  &        1.198     &    -0.213  &         0.832        &       -2.603    &        2.094     \\
$\phi_3$  &       0.3794  &        0.513     &     0.740  &         0.460        &       -0.626    &        1.385     \\
$\theta_1$  &       0.7441  &        0.946     &     0.787  &         0.432        &       -1.110    &        2.598     \\
$\theta_2$  &       0.4544  &        1.186     &     0.383  &         0.702        &       -1.870    &        2.779     \\
$\sigma^2$ &       0.0057  &        0.002     &     2.366  &         0.018        &        0.001    &        0.010     \\
\bottomrule
\end{tabular}
\begin{tabular}{lclc}
\textbf{Ljung-Box ($\phi_1$) (Q):}     & 0.09 & \textbf{  Jarque-Bera (JB):  } & 50.47  \\
\textbf{Prob(Q):}                & 0.76 & \textbf{  Prob(JB):          } &  0.00  \\
\textbf{Heteroskedasticity (H):} & 2.61 & \textbf{  Skew:              } & -2.02  \\
\textbf{Prob(H) (two-sided):}    & 0.17 & \textbf{  Kurtosis:          } &  8.20  \\
\bottomrule
\end{tabular}
\caption{Results of the ARMA model describing the dynamics of new retweeters of Musk.}
\end{table}